\documentclass[runningheads]{llncs}
\usepackage{changepage}
\usepackage{float}
\usepackage{cite}
\usepackage{multirow}
\usepackage{makecell}
\usepackage{booktabs}
\usepackage{hyperref}
\usepackage[T1]{fontenc}
\usepackage[utf8]{inputenc}

\usepackage{graphicx,verbatim}

\renewcommand{\figurename}{Figure}
\begin{document}
\title{Topology-Driven Fusion of nnU-Net and MedNeXt for Accurate Brain Tumor Segmentation on Sub-Saharan Africa Dataset}
\author{
Prabin Bohara*\inst{1}\and
Pralhad Kumar Shrestha\inst{2}\and
Arpan Rai\inst{3} \and
Usha Poudel Lamgade\inst{4} \and
Confidence Raymond\inst{5,6} \and
Dong Zhang\inst{5,7} \and
Aondona Lorumbu\inst{8} \and
Craig Jones\inst{9,10} \and
Mahesh Shakya\inst{11} \and
Bishesh Khanal\inst{11} \and
Pratibha Kulung\inst{12}
}

\authorrunning{Bohara et al.} % Use first author last name or "et al." if many
\titlerunning{Topology-Driven Fusion method for Brain Tumor Segmentation}

\institute{
Institute of Engineering, Thapathali Campus, Nepal\and
Gandaki College of Engineering and Science, Pokhara University, Pokhara, Nepal\and
Nepal Engineering College, Changunarayan-4, Bhaktapur, Nepal \and
Madan Bhandari University of Science and Technology, Chitlang, Nepal \and
Montreal Neurological Institute, McGill University, Montreal, QC, Canada \and
Department of Biomedical Engineering, McGill University \and
Medical Artificial Intelligence Laboratory (MAI Lab), Lagos, Nigeria \and
Department of Physics, Federal University of Technology, Minna, Nigeria \and
Department of Computer Science, Johns Hopkins University \and Department of Radiology and Radiological Science, Johns Hopkins School of Medicine \and
Nepal Applied Mathematics and Informatics Institute for Research (NAAMII), Nepal \and
Institute of Engineering, Purbanchal Campus, Nepal \\
\email{\{prabinbohara10, pralhad.shrestha05, mail.arpanrai, poudelusha7, mosesiorumbur, pratibha.kulu63\}@gmail.com} \\
\email{confidence.raymond@mail.mcgill.ca, donzhang@ece.ubc.ca} \\
\email{craig@imagingai.org}, \email{\{bishesh.khanal, mahesh.shakya\}@naamii.org.np}.
}

\maketitle
\begin{abstract}
Accurate automatic brain tumor segmentation in Low
and Middle-Income (LMIC) countries is challenging due to the lack of defined national imaging protocols, diverse imaging data, extensive use of low-field Magnetic Resonance Imaging (MRI) scanners and limited health-care resources. As part of the Brain Tumor Segmentation (BraTS) Africa 2025 Challenge, we applied topology refinement to the state-of-the-art segmentation models like nnU-Net, MedNeXt, and a combination of both. Since the BraTS-Africa dataset has low MRI image quality, we incorporated the BraTS 2025 challenge data of pre-treatment adult glioma (Task 1) to pre-train the segmentation model and use it to fine-tune on the BraTS-Africa dataset. We added an extra topology refinement module to address the issue of deformation in prediction that arose due to topological error. With the introduction of this module, we achieved a better Normalized Surface Distance (NSD) of 0.810, 0.829, and 0.895 on Surrounding Non-Enhancing FLAIR Hyperintensity (SNFH) , Non-Enhancing Tumor Core (NETC) and Enhancing tumor (ET).


\end{abstract}

\section{Introduction}
Gliomas are aggressive and life-threatening brain tumors, originating from glial cells in the brain or the spinal cord \cite{ref_chandana2008, ref_iranmehr2023}. They account for approximately 80 percent of glioma patients dying within two years of diagnosis. Despite advancements in diagnosis and treatment in high-income countries (HICs), mortality in Low and Middle-Income Countries (LMICs) like Sub-Saharan Africa (SSA) continues to rise high \cite{ref_patel2019, ref_poon2020, ref_adhikari2024}. This is mainly due to restricted access to medical infrastructure and trained professionals, urban-centric radiological talent bottlenecks, and delayed diagnosis \cite{bajwa2024scoping, cazap2016structural, brand2019delays}.

Magnetic Resonance Imaging (MRI) remains the clinical standard modality for detecting gliomas \cite{ref_verburg2020, ref_martucci2023}. It provides comprehensive visualization of the tumor and how far it has progressed using various imaging sequences. Still, in real-world practice, diagnosing and analyzing these tumors relies on skilled professionals who need to painstakingly draw the tumor borders scan by scan. This method is subjective and relies heavily on the peculiarities of a given radiologist’s skill and experience, making it very arduous both in terms of time and labor. As patient volume grows, manual segmentation becomes increasingly infeasible, delaying treatment and compromising outcomes.

The problem gives a clear indication of the need for automated solutions. 
The automatic segmentation of brain tumors is essential for the accurate and efficient planning
of surgery, evaluating therapeutic response, and determining the progression or recurrence of tumors \cite{ref_batool2024}.
Automation can alleviate clinician burden, increase diagnostic throughput, and improve workflow efficiencies, particularly in settings with limited resources and staff relative to workload.

The Brain Tumor Segmentation (BraTS) Challenge has acted as a benchmark test to build and assess artificial intelligence (AI) systems that segment brain tumors on MRI scans. The effectiveness of these models in resource-poor regions, such as SSA where MRI machines have low field strength and reduced contrast, poses a vital concern.  In this regard, the creation of the MICCAI BraTS-Africa dataset stands out as a major innovation. The dataset represents images acquired from imaging systems in SSA, characterized by reduced contrast and high artifacts, reflecting imaging conditions in resource-constrained regions. These limitations highlight the need for an architecture like U-Net, which can perform relatively well on low-field MRI data.

U-Net architecture is a mainstay in medical image segmentation because of its effectiveness, many open-source implementations, and consistently dependable performance across a variety of datasets \cite{ref_ronneberger2015, ref_dorfner2025}. However, its efficacy on novel or difficult datasets may be limited due to its inflexible structure and manual hyperparameter setup. On the other hand, nnU-Net performs exceptionally well by automatically modifying its network architecture, training parameters, postprocessing tactics, and preprocessing methods to fit the unique features of every dataset \cite{ref_magadza2023, ref_isensee2024}. Hence, it is particularly effective at complicated tasks like brain tumor segmentation.

On that account, we chose nnU-Net as our baseline model for the BraTS Africa Challenge. While nnU-Net performs better than the original U-Net on its own, we further improved performance by ensembling it with MedNeXt \cite{ref_hashmi2025, ref_liu2025}. However, both models still tend to optimize for pixel-to-pixel overlap, which can result in smoothed or fragmented segmentation of fine details, such as thin structures and small edges. Our ensemble includes a universal topology refinement \cite{ref_adewole2025, li2024universal } module to tackle this issue.

\section{Methods}
\subsection{Data}
The BraTS-Africa Challenge has the dataset of pre-operative glioma cases in African adults. It includes volumetric images from multiple scanners, including T1-weighted (T1), T2-weighted (T2), post-contrast T1-weighted (T1c), and T2 Fluid-Attenuated Inversion Recovery (T2-FLAIR). All the scans were pre-processed, manually annotated and segmented according to the standardized BraTS protocols \cite{ref_bratswiki, ref_adewole2023,ref_adewole2025}. These segmentations highlight three key tumor subregions: Enhancing Tumor (ET), Non-Enhancing Tumor Core (NETC), Surrounding Non-Enhancing FLAIR Hyperintensity (SNFH) or edematous region. For the BraTS Africa Challenge 2025, 60 training cases and 35 validation cases with the corresponding tumor sub-region masks from the BraTS-Africa 2024 were used to train and validate our proposed model.\cite{ref_bratswiki, ref_adewole2023,ref_adewole2025}.

\addtocounter{footnote}{-1}
\renewcommand{\thefootnote}{}
\footnotetext{The code for this paper is available at \url{https://github.com/SPARK-Academy-2025/SPARK-2025/tree/main/SPARK2025_BraTs_MODELS/Team_Saipal}}
\renewcommand{\thefootnote}{\arabic{footnote}}

\subsection{Topology Aware Segmentation}
Deep learning architectures are widely used in medical image segmentation, including the BraTS Challenge \cite{bonato2025advancing}. Even these sophisticated architectures fail to achieve pixel-wise accuracy due to the variance of resolutions of the images \cite{berger2024pitfalls, bohlender2021survey}. The need for topological correctness of image segmentation has led to the development of various metrics, including the lesion-wise dice score, lesion-wise Hausdorff distance-95 (HD95), and Normalized Surface Distance (NSD) \cite{Saluja2023BraTS2023Metrics, berger2024pitfalls, reinke2021common}. NSD@1 counts surface points within a distance of 1 (in voxels or mm) as a correct prediction. Among them, legacy metrics compute the score based on the entire brain volume, and lesion-wise metrics compute scores by giving equal importance to each tumor lesion regardless of its size \cite{ren2024optimization}.

\begin{figure}[h]
\includegraphics[width=\textwidth]{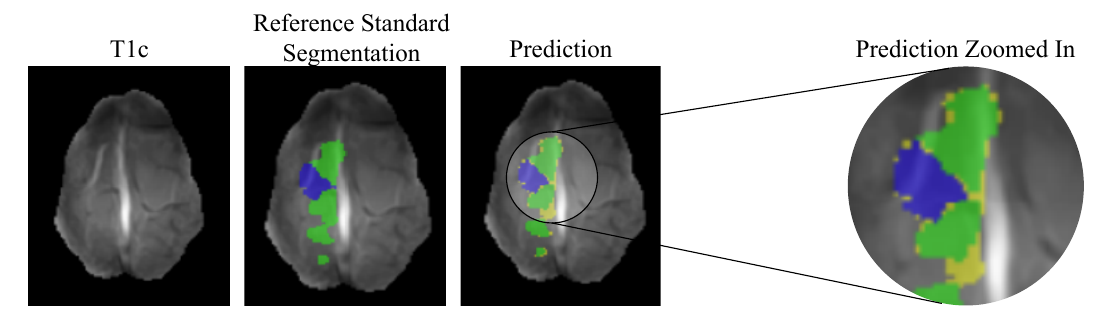}
\caption{Topological error visualization of reference standard segmentation and predicted output before topology refinement, where yellow color denotes prediction error.} \label{fig_topology_error}
\end{figure}

Figure \ref{fig_topology_error} shows a topological error in the segmentation predicted by nnU-Net on the SSA dataset, before the topology refinement.
It illustrates the comparison of the expert annotated segmentation and the predicted output from our baseline nnU-Net model, where the model struggles to predict the boundary areas. This could explain why our lesion-wise dice score is lower than the legacy dice score.

In order to correct these topological errors, most of the researchers have developed either their evaluation metrics or their segmentation techniques \cite{berger2024pitfalls, NEURIPS2019_2d95666e, NEURIPS2023_19ded4cf}. These approaches might only work for the specific task, as these approaches could lead to poor generalization when applied to different segmentation tasks. To address this challenge, Liu et al. proposed a universal topology preservation and refinement method  \cite{li2024universal}. This method creates topology-perturbation masks using randomly sampled coefficients of orthogonal polynomial bases that generate the unbiased representation of the predictions, and promotes better generalization.
\subsection{Model}
We built and trained networks using various configurations of nnU-Net, MedNeXt, ensemble, and topology refinement methods, incorporating  pre-processing and exploring different data augmentation techniques. We trained both two-dimensional (2D) neural networks on a slice based MRI data and three-dimensional (3D) full-resolution data for 500 epochs. This configuration was used both for the vanilla BraTS-Africa dataset as well as the pretraining approach on the 2025 BraTS Glioma Dataset. We further refined the segmentation using a post topological refinement via polynomial feature synthesis \cite{li2024universal}.

\subsubsection{Baseline nnU-Net :} We trained 2D and 3D full-resolution nnUnet on the voxel size of 128 × 160 × 112 with deep supervision. The architecture has six encoder-decoder levels with 32, 64, 128, 256, 320, and 320 features, respectively. The convolution kernel sizes (3 × 3 × 3) were uniform across the architecture and were performed twice per level. Due to default batch size being set to 2, batch normalization was applied instead of Instance Normalization.

\subsubsection{Baseline MedNeXt :} MedNeXt based on ConvNeXt and 3D U-Net was used as our second baseline model. It has six residual stages with depthwise convolution with a kernel size of 3 × 3 × 3, channel-wise Group Normalization and GELU activation with and deep supervision at decoder stages to stabilize training and improve gradient flow. The softmax output from both the baselines would be combined by soft voting averaging.

We trained our method on an NVIDIA V100 GPU, each training epoch required approximately 80 seconds, with the total training completing in about 11.1 hours.

\begin{figure}[h]
\includegraphics[width=\textwidth]{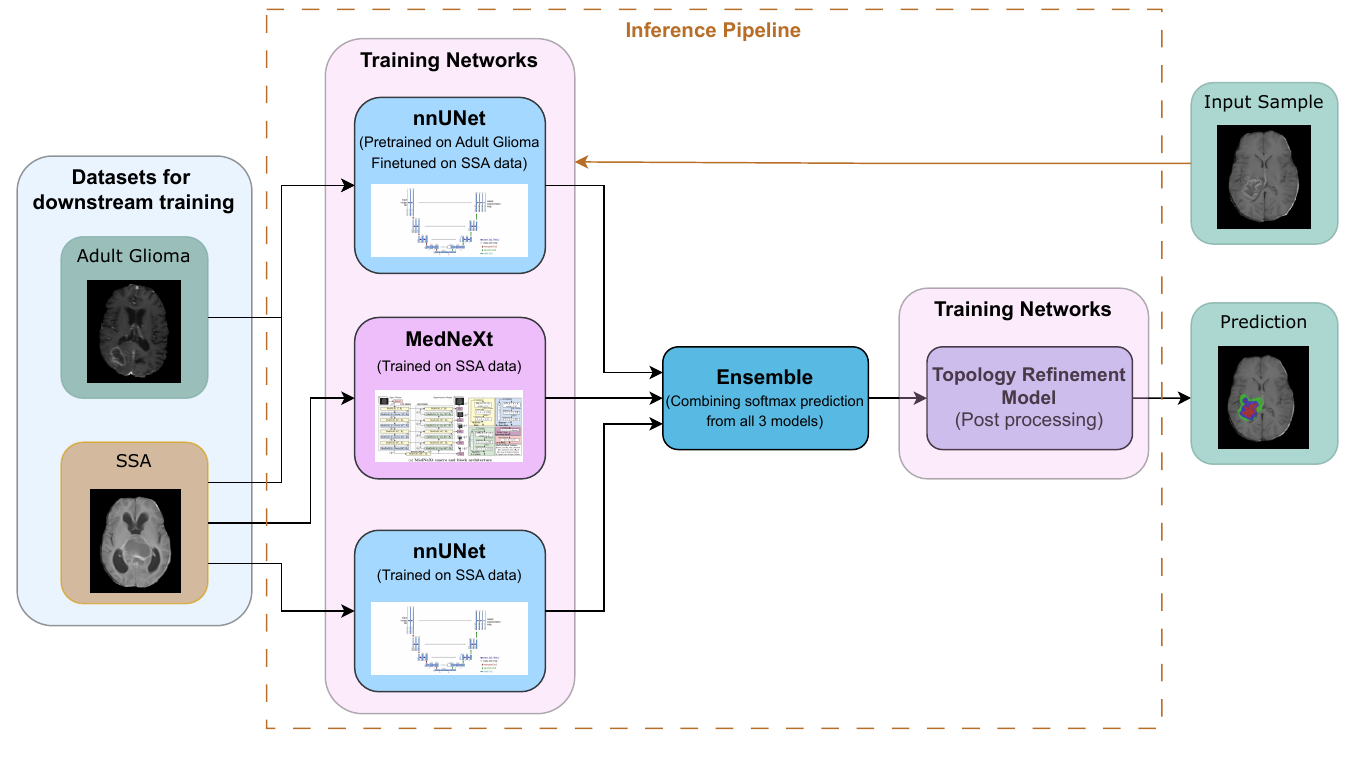}
\caption{Model of our proposed Pipeline using an Ensemble of nnU-Net and MedNeXt with Topology-aware Post-Processing} \label{fig1}
\end{figure}

\subsubsection{Topology Refinement Model :} We adopted the Universal Topology Refinement approach to address topological error correction \cite{li2024universal}, which generated synthetic segmentation labels and also provided a trainable pipeline network \cite{ref_li2023lncs} to reduce topological errors in the baseline model. We used the topology-perturbation mask to introduce structural errors into synthetic data that imitated the common segmentation model errors, and used it to train our post-processing topology refinement network. Our network used a 3D U-Net of size $240 \times 240 \times 155$, and we experimented with 4 to 8 input feature maps to generate the 4 output feature maps in the final layer. This model was used to detect and fix the topological errors that our baseline or ensemble model missed. The results of our baseline model were used to generate the final prediction of this model.

\section{Results}
\begin{table}
\centering
  \renewcommand{\arraystretch}{1.2}
  \setlength{\tabcolsep}{2.5pt}
  
  \caption{Performance Dice Similarity Coefficient (DSC) of different models on Legacy and Lesion regions. Best results are in \textbf{bold}.}
  
  \begin{tabular}{|l|ccc|ccc|}
    \hline
     & \multicolumn{3}{c|}{\textbf{Legacy (DSC)}} & \multicolumn{3}{c|}{\textbf{Lesion (DSC)}} \\
     \cline{1-7}
    {Model} & \ SNFH & \ NETC & \ ET & \ SNFH & \ NETC & \ ET \\
    \hline
    nnU-Net(3D full-res) & 0.930 & 0.906 & 0.906  & 0.891 & \textbf{0.853} & \textbf{0.856}\\
    nnU-Net(2D) & 0.915 & 0.847 & 0.852  & 0.719 & 0.720 & 0.751  \\
    MedNeXt & 0.920 & 0.891 & 0.891  & 0.902 & 0.846 & 0.848  \\
    Topology-aware & 0.920 & 0.901 & 0.900 & 0.853 & 0.845 & 0.848 \\
    Fine-tuned on BraTS-Africa & \textbf{0.936} & \textbf{0.907} & 
\textbf{0.907} & 0.896 & 0.848 & 0.853 \\
    Ensemble & 0.934 & 0.878 & 0.880 & \textbf{0.908} & 0.824 & 0.839 \\
    \hline
  \end{tabular}
  \label{tab:DSC_comparison}
\end{table}

Our Baseline Model, nnU-Net 3D full resolution performed well overall on the BraTS-Africa dataset achieving high Dice scores for each tumor subregion: 0.930 for SNFH, 0.906 for NETC, and 0.906 for ET.
When pre-trained with the BraTS-2025 Task 1 (adult glioma-pre-treatment) dataset and fine-tuned with the BraTS-Africa dataset, there is a slight increase in the Legacy dice score, achieving 0.936 for SNFH, 0.907 for NETC, and 0.907 for ET.

The boundary alignment was demonstrated on NSD with a tolerance of 1.0 mm, for SNFH (0.830), NETC (0.827), and ET (0.894).
The MedNeXt, fine-tuned, and ensemble models achieved Dice and NSD scores that were comparable to the baseline. In contrast, the topology-aware refinement model did not outperform the baseline in terms of Dice score but yielded a subtle improvement in NSD. Specifically, this model achieved Dice scores of SNFH (0.920), NETC (0.901) and ET (0.900) with corresponding NSD values of 0.810, 0.829, and 0.895, respectively. While a slight improvement in NSD was observed for the NETC and ET regions, a decline in SNFH score was observed, suggesting a trade-off between topological consistency and overall surface agreement in certain subregions.

\section{Discussion}
Our baseline model was trained on both 2D patches and 3D full-resolution data, accommodating clinical setups that often rely on 2D image acquisition. Our experiments illustrated the similarity and contrast of the segmentation outputs using different sets of trained and fine-tuned models. Taking inspiration from Li et al. and extending Isensee et.al. work, we introduced an additional universal topology refinement module applied to the ensemble outputs. This post-processing step enhanced boundary precision and better captures thin structures and sharp edges within tumor subregions.

\begin{table}
\centering
  \renewcommand{\arraystretch}{1.2}
  \setlength{\tabcolsep}{2.5pt}
    
  \caption{Performance (NSD at 0.5) of different models on Legacy and Lesion regions. Best results are in \textbf{bold}.}
  
  \begin{tabular}{|l|ccc|ccc|}
    \hline
     & \multicolumn{3}{c|}{\textbf{Legacy (NSD\textnormal{@}0.5)}} & \multicolumn{3}{c|}{\textbf{Lesion (NSD\textnormal{@}0.5)}} \\
     \cline{1-7}
    {Model} & \ SNFH & \ NETC & \ ET & \ SNFH & \ NETC & \ ET \\
    \hline
    nnU-Net(3D full-res) & 0.539 & 0.537 & 0.605  & 0.517 & 0.514 & 0.575\\
    
    nnU-Net(2D) & 0.428 & 0.423 & 0.498  & 0.340 & 0.359 & 0.434  \\
    
    MedNeXt & 0.535 & \textbf{0.543} & \ 0.601  & 0.523 & \textbf{0.522} & 0.574  \\
    
    Topology-aware & \ 0.518 & 0.541 & \textbf{0.606} & \ 0.484 & 0.520 & \textbf{0.579} \\
    Fine-tuned on BraTS-Africa & 0.546 & 0.528 & 0.600 & 0.523 & 0.502 & 0.568 \\
    Ensemble & \textbf{0.572} & 0.533 & 0.600 & \textbf{0.554} & 0.509 & 0.574 \\
    \hline
  \end{tabular}
  \label{tab:NSD@0.5_comparison}
\end{table}

\begin{table}
\centering
  \renewcommand{\arraystretch}{1.2}
  \setlength{\tabcolsep}{2.5pt}
  
  \caption{Performance (NSD at 1.0) of different models on Legacy and Lesion regions. Best results are in \textbf{bold}.}
  
  \begin{tabular}{|l|ccc|ccc|}
    \hline
     & \multicolumn{3}{c|}{\textbf{Legacy (NSD\textnormal{@}1.0)}} & \multicolumn{3}{c|}{\textbf{Lesion (NSD\textnormal{@}1.0)}} \\
     \cline{1-7}
    {Model} & \ SNFH & \ NETC & \ ET & \ SNFH & \ NETC & \ ET \\
    \hline
    nnU-Net(3D full-res) & 0.830 & 0.827 & 0.894  & 0.796 & 0.787 & 0.849\\
    
    nnU-Net(2D) & 0.756 & 0.713 & 0.810  & 0.603 & 0.609 & 0.713  \\
    
    MedNeXt & 0.825 & \ 0.812 & \ 0.875  & 0.809 & \ 0.778 & 0.836  \\
    
    Topology-aware & \ 0.810 & \textbf{0.829} & \textbf{0.895} & \ 0.755 & \textbf{0.790} & \textbf{0.850} \\
    Fine-tuned on BraTS-Africa & 0.840 & 0.821 & 0.893 & 0.805 & 0.776 & 0.842 \\
    Ensemble & \textbf{0.850} & 0.805 & 0.873 & \textbf{0.826} & 0.764 & 0.835 \\
    \hline
  \end{tabular}
  \label{tab:NSD@1.0_comparison}
\end{table}

\begin{figure}
\includegraphics[width=\textwidth]{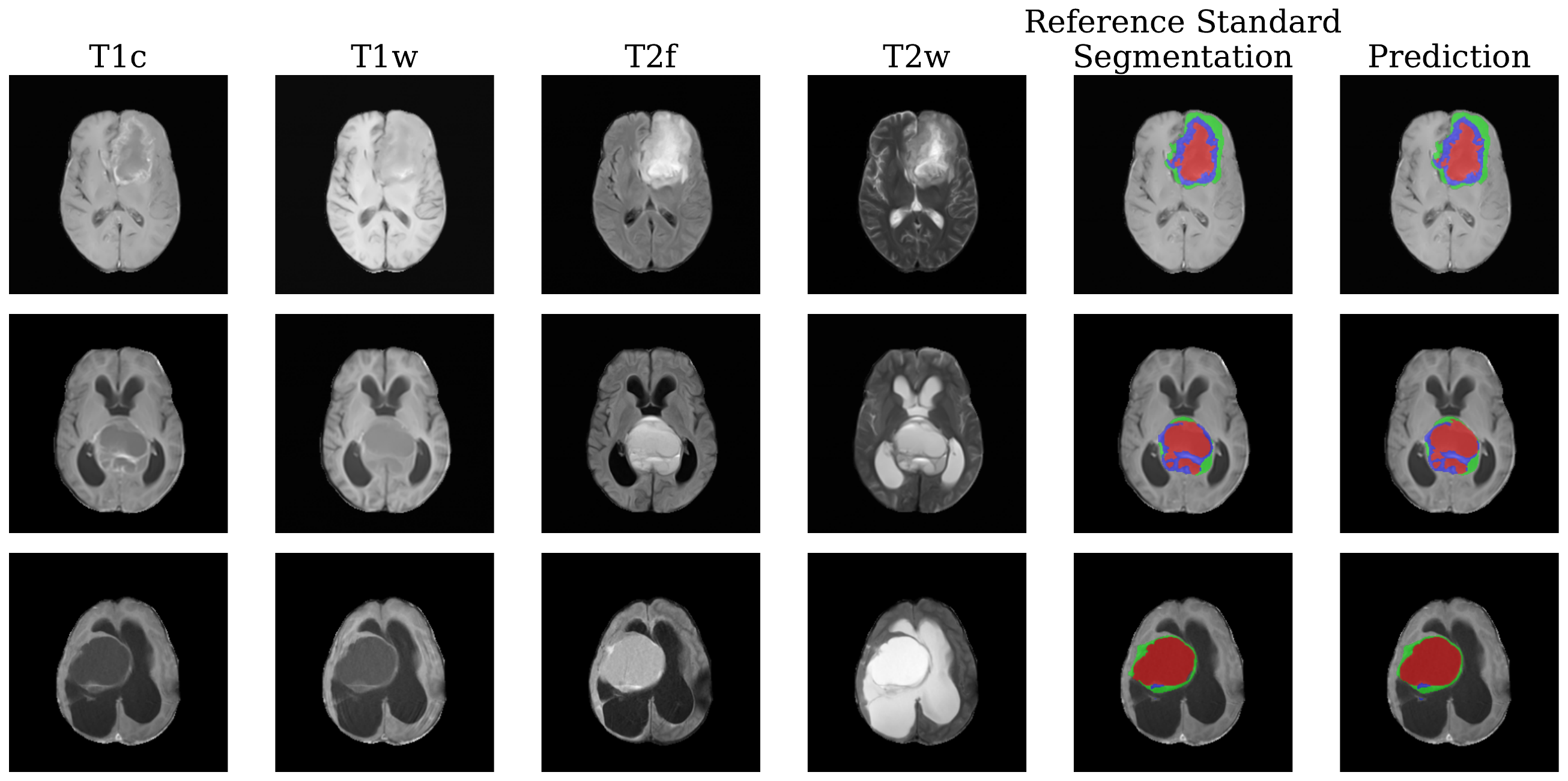}
\caption{Visualization of multimodal MRI inputs (T1c, T1w, T2-FLAIR, T2w) for cases BraTS-SSA-00010-000, BraTS-SSA-00025-000, BraTS-SSA-00096-000, along with corresponding reference standard segmentation masks, and baseline nnU-Net predicted tumor segmentation results. The segmentation overlays highlight tumor subregions including non-enhancing tumor (red), enhancing tumor (blue), and surrounding non-enhancing FLAIR hyperintensity (green). The predictions closely match the reference standard annotations, demonstrating the model’s effectiveness.} \label{figure.2.}
\end{figure}

It demonstrated comparatively better region-wise performance for the NETC and ET in both Dice and NSD scores. Although the fine-tuned model on SSA achieved the highest Legacy Dice score, it showed a drop in lesion-wise Dice scores and NSD performance. Since lesion-wise performance is more accurate for evaluating the segmentation quality \cite{Saluja2023BraTS2023Metrics}, the fine-tuned model was not selected as the baseline for training the topology refinement model. The trained topology refinement model resulted in only minimal improvements in NSD for the NETC and ET regions. However, it showed a drop in Dice scores, indicating a potential trade-off between surface-level refinement and volumetric segmentation accuracy.

The post-processing phase was implemented with the aim of improving boundary precision and more effectively capturing thin structures and sharp edges in tumor subregions. Unexpectedly, while the topology refinement module enhanced boundary delineation, as illustrated in Figure \ref{fig_topology_error}, it failed to produce better Dice scores compared to the baseline model. This could be due to the refinement technique being more successful in maintaining vascular structures instead of tumor characteristics, or further architectural enhancements might be necessary to improve the method for brain tumor segmentation. Therefore, more research on refining model design and enhancement techniques is needed to get the best performance from topology-aware methods in brain tumor segmentation tasks.

\section{Conclusion}
Our study explored the BraTS-Africa dataset using a multi-model technique to improve brain tumor segmentation. We used nnU-Net as the baseline segmentation model, paired it with MedNeXt, and incorporated a topology refinement module to improve the combined predictions even further. Although the topology refinement approach resulted in only marginal improvements in our study, we believe it remains a promising direction for addressing pixel-wise accuracy and achieving topologically correct image segmentation. We truly believe that this study can be extended further to achieve a state-of-the-art segmentation model through the combined capability of self-configuring baseline models and the enhancement through topology corrections. Future work should explore optimized topology-aware loss functions to further improve segmentation performance.

\section{Conflict of Interest}
The authors declare that there are no conflicts of interest. 

\section{Acknowledgement}
The authors would like to thank the instructors of the Sprint AI Training for African Medical Imaging Knowledge Translation (SPARK) Academy 2025 summer school. The authors acknowledge the computational infrastructure support from the Digital Research Alliance of Canada (The Alliance) and the University of Washington Azure GenAI for Science Hub through The eScience Institute and Microsoft (PI: Mehmet Kurt) secured the SPARK Academy. Finally, we would like to thank the Lacuna Fund for Health and Equity (PI: Udunna Anazoda), the Radiological Society of North America (RSNA), the Research \& Education (R\&E) Foundation Derek Harwood-Nash International Education Scholar Grant (PI: Farouk Dako), the McGill University Healthy Brain and Healthy Lives (HBHL; Udunna Anazoda) and the National Science and Engineering Research Council of Canada (NSERC) Discovery Launch Supplement (PI: Udunna Anazoda) for making the SPARK Academy possible via research grant supports.
Additionally, we would like to sincerely thank the Nepal Applied Mathematics and Informatics Institute for Research (NAAMII) for their unwavering support and guidance during this project.

\newpage
\bibliographystyle{splncs04}
\bibliography{bibliography}

\newpage
\appendix
\section*{Supplementary Materials}
\renewcommand{\thesubsection}{S\arabic{subsection}}

\begin{figure}[h]
\includegraphics[width=\textwidth]{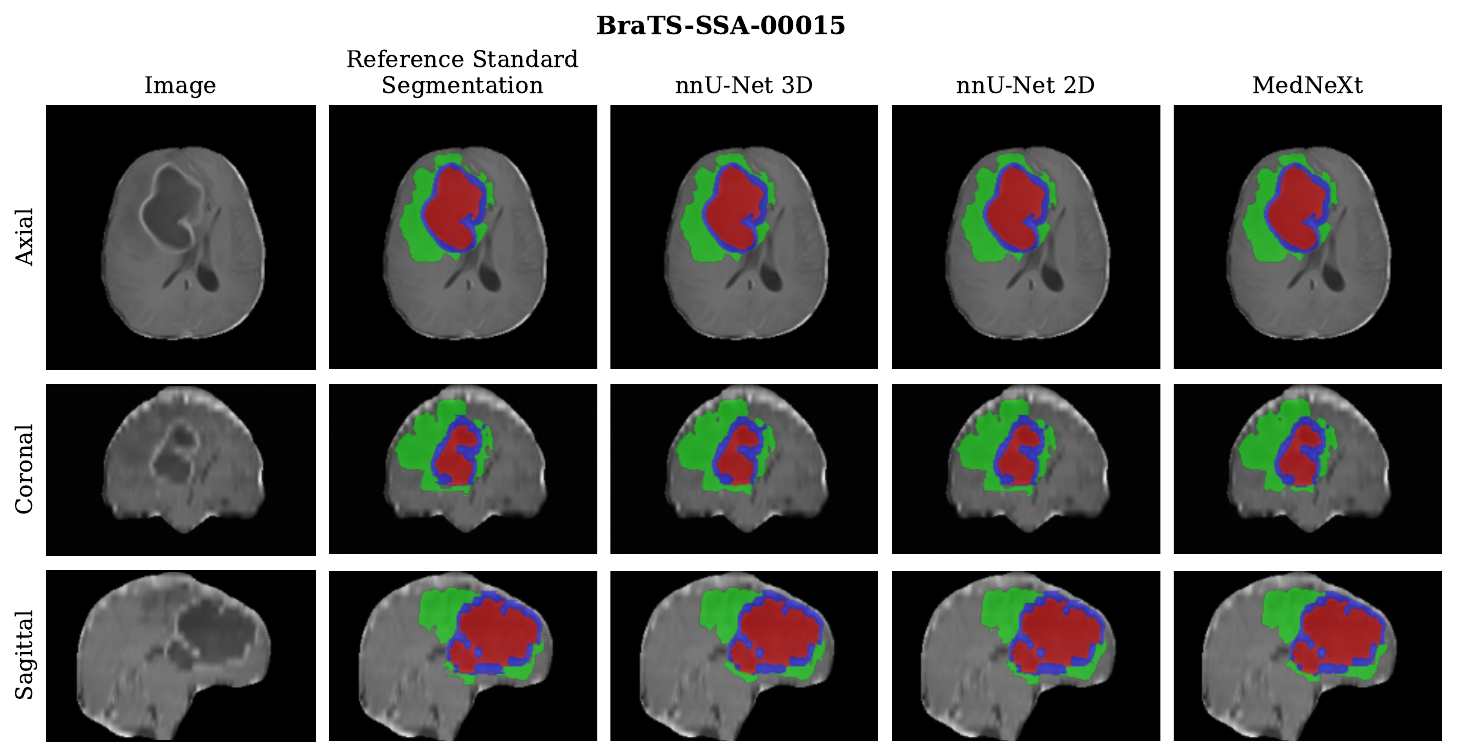}
\caption{Multiplanar visualization (Axial, Coronal, and Sagittal views) of a brain MRI scan from the \textbf{BraTS-Africa-00015} case. The first column shows the original T1c MRI slices. The second column overlays the ground truth tumor segmentation mask on the MRI, while the third, fourth and fifth columns overlay the predicted segmentation masks from the nnU-Net 3D, nnU-Net 2D, and MedNeXt models, respectively. This layout provides a side-by-side visual comparison of each model's predictions directly on the anatomical scan. The segmentation overlays highlight tumor subregions including non-enhancing tumor core (red), enhancing tumor (blue), and surrounding non-enhancing FLAIR hyperintensity (green).} \label{figure.3.}
\end{figure}

\begin{table}[H]
\centering
  \renewcommand{\arraystretch}{1.2}
  \setlength{\tabcolsep}{2.5pt}
  
  \caption{Topology-aware model performance on lesion regions (DSC and NSD@1.0) from the competition testing phase.}
  
  \begin{tabular}{|l|ccc|ccc|}
    \hline
     & \multicolumn{3}{c|}{\textbf{Lesion (DSC)}} & \multicolumn{3}{c|}{\textbf{Lesion (NSD@1.0)}} \\
     \cline{1-7}
    {Model} & \ SNFH & \ NETC & \ ET & \ SNFH & \ NETC & \ ET \\
    \hline
    Topology-aware & 0.874 & 0.845 & 0.836  & 0.804 & 0.803 & 0.837\\
    \hline
  \end{tabular}
  \label{tab:testing_results}
\end{table}

\end{document}